\documentclass[pra,twocolumn,floatfix]{revtex4}
\usepackage{graphicx}
\usepackage{amsmath}
\newcommand{\up}{\uparrow}
\newcommand{\down}{\downarrow}
\begin{document}

\title{Density correlations in ultracold atomic Fermi gases}
\author{W. Belzig}
\affiliation{University of Konstanz, Department of Physics, 
D-78457 Konstanz, Germany}
\author{C. Schroll}
\affiliation{Department of Physics and Astronomy, University of Basel,
 Klingelbergstr. 82, 4056 Basel, Switzerland} 
\author{C. Bruder}
\affiliation{Department of Physics and Astronomy, University of Basel,
 Klingelbergstr. 82, 4056 Basel, Switzerland} 
\date{\today}

\begin{abstract}
We investigate density fluctuations in a coherent ensemble of
interacting fermionic atoms. Adapting the concept of full counting
statistics, well-known from quantum optics and mesoscopic electron
transport, we study second-order as well as higher-order correlators
of density fluctuations.  Using the mean-field BCS state to describe
the whole interval between the BCS limit and the BEC limit, we obtain
an exact expression for the cumulant-generating function of the
density fluctuations of an atomic cloud.  In the two-dimensional case,
we obtain a closed analytical expression.  Poissonian fluctuations of
a molecular condensate on the BEC side are strongly suppressed on the
BCS side. The size of the fluctuations in the BCS limit is a direct
measure of the pairing potential.  We also discuss the BEC-BCS
crossover of the third cumulant and the temperature dependence of the
second cumulant.
\end{abstract}



\maketitle

\section{Introduction}

Following the successful creation of Bose-Einstein condensates (BECs)
in ultracold atomic clouds \cite{bec:95}, recently ultracold Fermionic
clouds have been produced
\cite{DeMarco:99,Truscott:01,Schreck:01,Granade:02,OHara:02}.  This has
attracted a lot of 
attention both theoretically and experimentally, especially due to the
ability to tune the mutual interaction between atoms via a
Fano-Feshbach resonance.  The unique opportunity to study the
crossover from weak attractive to strong attractive interactions in
one and the same system makes this interesting from a fundamental
many-body point of view.

Theoretically, fermionic systems with weak attractive interaction are
superfluids and as such described by the Bardeen-Cooper-Schrieffer
(BCS) theory \cite{bcs:57}. This theory
can also describe the limit of stronger attractive interaction
\cite{eagles:69,leggett:80,NSR:85}, in which a BEC of molecules is
formed. In the crossover regime, the
long-range nature of the interaction makes the BCS theory less
accurate \cite{strinati:04}.

Recently, several experiments studied the strongly interacting BEC-BCS
crossover regime using spin mixtures of ultracold fermionic gases (see
\cite{grimm2007} for a recent review).  Measurements of the
interaction energy of an ultracold fermionic gas near a Feshbach
resonance were made, studying the impact of the interaction on the
time of flight expansion \cite{Bourdel:03}. Experimental investigation
of collective excitations showed a strong dependence on the coupling
strength \cite{Bartenstein:04,Kinast:04}. Moreover, the condensation
\cite{Regal:04,Zwierlein:04} and the spatial correlations
\cite{Greiner:04} of the fermionic atom pairs were observed in the
full crossover regime.  The pairing gap was measured directly via a
spectroscopic technique in the whole crossover region
\cite{denschlag:04}. Remarkably, the gap values are in relatively good
agreement with the BCS model in the whole region.

The use of noise correlations to probe the many-body states of
ultracold atoms was proposed in \cite{lukin:04} (see also
\cite{lamacraft05a}). Correlation measurements can be applied to detect
phase coherence in mesoscopic superpositions \cite{bach:04}.  The
density and spin structure factor for the BEC-BCS transition was
calculated \cite{buechler:04}.  Interferometric measurement schemes of
the spatial pairing order have been proposed based on the atom
counting statistics in the output channels \cite{Carusotto:04}.
Pairing fluctuations of trapped Fermi gases have been studied in
\cite{viverit:04}. The number statistics of Fermi and Bose gases has
also been investigated in \cite{budde:04,meiser:04}.  In recent
experiments \cite{grimm:04} the spatial structure of an atomic cloud
has been directly observed (without the expansion used in most other
experiments). This makes it possible to determine the density
fluctuations either by repeating the experiment many times or by
taking densities at different positions in a homogeneous system to
extract the statistics.  Atomic shot noise has been experimentally
investigated both in bosonic and fermionic systems
\cite{foelling:05,aspect2005,greiner:05,bloch2006,esteve2006,aspect2007,porto2007}.

In this article we propose to use the full counting statistics of
density fluctuations as a tool to gain access to the many-body nature
of the ground state of a fermionic cloud in the BEC-BCS crossover
regime (for other work on full counting statistics in ultracold atomic
gases see \cite{esslinger2005} on the experimental and
\cite{lamacraft05b,meystre,moritz,svistunov} on the theoretical
side). Our main result is a general expression for the particle number
statistics of the mean-field BCS wave function.
In the limiting cases, the statistics allows a straightforward
interpretation.  Deep in the molecular BEC limit the statistics is
Poissonian, i.e. that of independent pairs of atoms. In the opposite
limit, on the BCS side of the crossover, the fluctuations are strongly
suppressed and reflect the particle-hole symmetry.  The statistics in
the crossover regime differs strongly from both the BCS and BEC limits
and will be discussed based on several numerical results.

\section{Counting statistics of density fluctuations}

\begin{figure}[htbp]
  \centering
  \includegraphics[width=0.45\columnwidth]{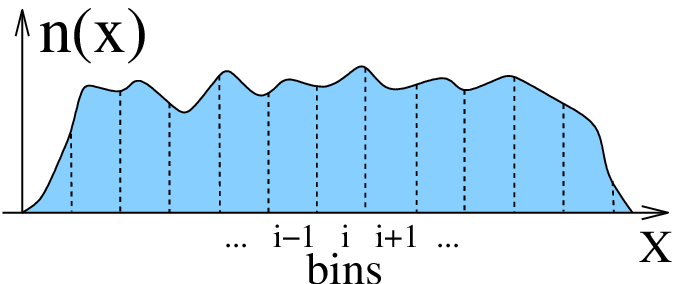}
  \includegraphics[width=0.45\columnwidth]{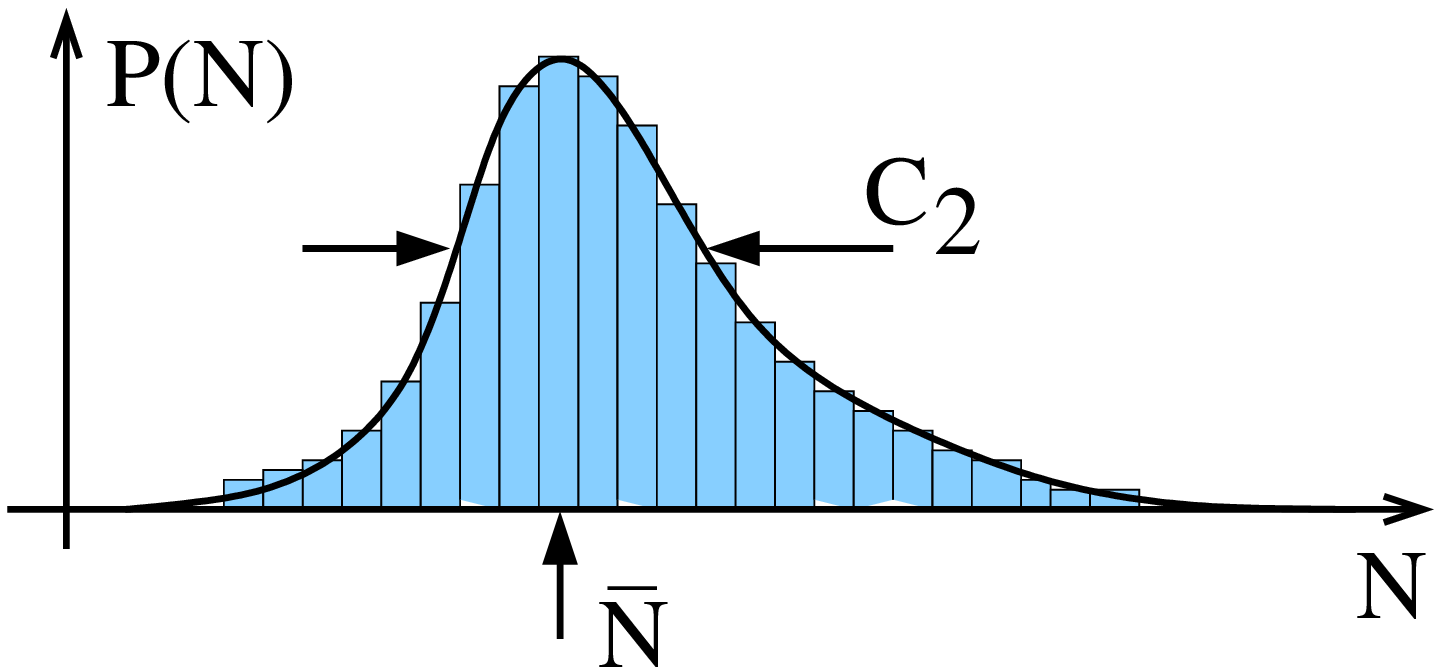}
  \caption{
(Color online)
(a) Sketch of a typical atomic number density. The measured observable is
 the atom number $N$ given by $n(x)$ integrated over the bin volume 
(the bins are indicated by the dashed lines).
(b) Histogram of probabilities to find $N$ particles in a bin.
}
  \label{fig:system}
\end{figure}

We consider an atomic cloud with spatial distribution $n(x)$ and
divide the system into bins, see Fig.~\ref{fig:system}.  We are
interested in the probability $P(N)=\langle \delta(N-\hat N)\rangle$
to find $N$ atoms in a ``bin'' of the system, see
Fig.~\ref{fig:system}. Here, $\hat N= \int_{V_{\mathrm{bin}}} \hat
n(x)$ is the bin number operator, $V_{\mathrm{bin}}$ is the bin volume,
and $\hat n(x)$ denotes the density operator. The bin volume is
assumed to be much smaller than the volume of the full system, but
still much larger than the mean interparticle distance cubed.  Hence,
a single bin can be considered as a grand-canonical system with the
surrounding atomic cloud serving as the reservoir.

In practice, it is more convenient to study the Fourier transform of
$P(N)$, which is related to the cumulant generating function (CGF)
$S(\chi)$ via
\begin{equation}
  \label{eq:cgf-definition}
  e^{-S(\chi)}=\sum_N e^{i\chi N} P(N) =\langle e^{i\chi \hat N} \rangle\,.
\end{equation}
The cumulants $C_n$ are defined in a standard way as $S(\chi)=-\sum_n
C_n(i\chi)^n/n!$ and can be used to characterize the full counting
statistics (FCS).  We recall that $C_1=\bar N$ is the average number
of atoms in a bin, and $C_2=\langle (N-\bar N)^2\rangle$ measures the
width of the number distribution. The third cumulant is proportional
to the skewness and, therefore, a measure of the asymmetry of the
distribution function. As we will see later, the third cumulant in a
fermionic system is related to deviations from particle-hole symmetry.

\section{Noninteracting Fermions and Bosons}
\label{sec:non}

We start by recalling some properties of noninteracting Fermi and Bose
gases at a given temperature $T$. Here, individual atoms are
independent and we can obtain the statistics as
\begin{equation}
  \label{eq:cgf-fermibose}
  S(\chi)=\mp \sum_k \ln\left[1\pm f_\pm(k)(e^{i\chi}-1)\right]\, ,
\end{equation}
where $f_\pm(k)=\left[\exp((\epsilon_k-\mu)/k_BT)\pm 1\right]^{-1}$ is
the Fermi (Bose) occupation function. The chemical potential is
determined by the average atom number, i.~e., $C_1=\bar N=\sum_k
f_\pm$. Obviously, the number statistics of fermions and bosons
differs drastically in the degenerate regime. In the nondegenerate
regime $f_\pm$ is small and we find the Poissonian statistics of
classical particles $S(\chi)=-\bar N(\exp(i\chi)-1)$, both for
fermions and bosons.

For degenerate fermions $k_BT\ll \epsilon_F$ the statistics is 
\begin{equation}
  \label{eq:fermion}
  S(\chi)=-i\chi\bar N - (D k_BT/4 \epsilon_F)\bar N \chi^2  \,.
\end{equation}
Here and in the following, $S(\chi)$ is defined in the interval
$[-\pi,\pi]$ and extended periodically.  Thus, particle-number fluctuations are
suppressed by $T/\epsilon_F$ in comparison to the classical case
\cite{castin}. Remarkably, the statistics is Gaussian and consequently
all cumulants $C_n$ for $n \ge 3$ vanish.  This behavior resembles
fermions in a 1D wire \cite{levitov:96} and can be interpreted as a
consequence of anti-bunching. Note, that the Gaussian nature of the
statistics is a consequence of particle-hole symmetry and is therefore
strictly limited to the degenerate regime. Higher-order corrections in
$k_BT/\epsilon_F$ will introduce deviations from the Gaussian limit,
and lead e.g. to the appearance of higher-order odd cumulants, which
are directly related to deviations from the perfect particle-hole
symmetry.

In contrast to that, we obtain quite a different behavior for free
bosons.  Approaching the degeneracy temperature $T^{\mathrm{BEC}}_C
=2\pi\hbar^2 n^{2/3}/m\zeta(3/2)^{2/3}$ \cite{T_BEC} from above, the
fluctuations are enhanced due to the large factor $f_-(k)(f_-(k)+1)$.
In the condensed regime, the occupation of the ground state becomes
macroscopically large and the grand-canonical approach is no longer
valid \cite{landau,leggett2006}. 
This can be seen if we take the limit $T=0$ of
Eq.~(\ref{eq:cgf-fermibose}), leading to $S(\chi)=\ln\left[1-\bar
  N(e^{i\chi}-1)\right]$, which corresponds to a negative binomial
distribution and the fluctuations therefore diverge according to
$C_n\sim\bar N^n$.  As in the grand-canonical ensemble the chemical
potential is $\mu=0$ below the critical temperature
$T<T^{\mathrm{BEC}}_C$, an arbitrarily large number of bosons can be
transferred from the reservoir into the bin, leading to unphysically
large fluctuations.  In order to find the correct fluctuations in this
case, we calculate the FCS from Eq.~(\ref{eq:cgf-definition})
explicitly.  We divide the boson operator into $a_k = b_k + c_k$,
where
\begin{equation}
b_k=\int_{V_{\mathrm{bin}}} d^3r\;e^{ikr}\Psi(r)\,, \quad
c_k=\int_{V^\prime} d^3r\;e^{ikr}\Psi(r)\,,
\end{equation}
where $V^\prime$ is the volume $V$ without the volume
$V_{\mathrm{bin}}$ of the bin.  We consider a fully condensed state of
$N_{\mathrm{tot}}$ non-interacting bosons $|\psi\rangle =
(a_0^\dagger)^{N_{\mathrm{tot}}}|{\mathrm{vac}}\rangle = (b_0^\dagger
+ c_0^\dagger)^{N_{\mathrm{tot}}}|{\mathrm{vac}}\rangle$. Using
Eq.~(\ref{eq:cgf-definition}) for the bin number operator
$\hat{N} = \sum_k b_k^\dagger b_k$ we obtain a binomial
statistics for a non-interacting bosonic gas in a bin at $T=0$
\begin{equation}
\label{eq:cgf-binbosons}
S(\chi) = -N_{\mathrm{tot}}\ln{[1+\frac{V_{\mathrm{bin}}}{V}(e^{i\chi}-1)]}\,.
\end{equation}   
For bin volumes $V_{\mathrm{bin}}/V \ll 1$ the particle-number statistics
becomes Poissonian, i.e. $S(\chi)\approx
-\bar{N}(e^{i\chi}-1)$, where $\bar{N}=N_{\mathrm{tot}}V_{\mathrm{bin}}/V$.

\section{BCS ground state}

We now turn to a Fermi gas with an attractive interaction
parameterized by a scattering length $a$.  The Hamiltonian for the
full system is number conserving and as such shows no number
fluctuations at all. Here we consider a bin, i.e. a small subsystem,
which we assume to be described by the non-number conserving BCS
Hamiltonian \cite{bcs:57}. A straightforward ansatz for a non-number
conserving many-body state, which takes the pairing interaction into
account, is the BCS wave function, which will be used in the following
and is known to correctly describe both the BCS and the BEC limit
\cite{eagles:69,leggett:80,NSR:85}. We will later show that the
approach also reproduces the correct counting statistics in the two
limits, and therefore prefer to use this transparent, almost
analytical approach instead of more complex canonical approaches. 
The BCS wave function is given by
\begin{equation}
  \label{eq:bcs-wavefunction}
  \left|\mathrm{BCS}\right\rangle = \prod_k \left(u_k+v_k
  c_{k\up}^\dagger c_{-k\down}^\dagger \right)\left|0\right\rangle\,.
\end{equation}
The variational procedure yields $v_k^2 = 1-u_k^2 =
\left(1-(\epsilon_k-\mu)/E_k\right)/2$, where
$E_k^2=(\epsilon_k-\mu)^2+\Delta^2$ is the energy of 
quasiparticle excitations. The order parameter $\Delta$ and the chemical
potential are fixed by the self-consistency equations
 \begin{eqnarray}
\label{eq:gap-xi}
  \Delta=-\lambda\sum_k u_k v_k 
&,&
  \bar N = 2\sum_k v_k^2\,,
\end{eqnarray}
where $\lambda$ is the BCS coupling constant. After renormalization of
the coupling constant $\lambda$ and considering only the low-energy
limit, the gap equation can be related to the two-particle scattering
amplitude \cite{leggett:80,randeria:90}.

The product form of the BCS wave function greatly simplifies the
calculation of the statistics, since different $k$
states can be treated separately. For a single pair of states
$(k\up,-k\down)$ the sum over all possible configurations can be easily
performed
\begin{eqnarray}
  \label{eq:cgf-bcsk}
  e^{-S_k(\chi)}=
  \langle\mathrm{BCS}|e^{i(\hat n_{k\up}+\hat n_{-k\down})\chi}|
\mathrm{BCS}\rangle&&\\=u_k^2+v_k^2e^{2i\chi}\,.&&\nonumber
\end{eqnarray}
The sum of all states yields the result
\begin{equation}
  \label{eq:cgf-bcs}
  S(\chi)=-\sum_k \ln\left[1+v_k^2(e^{2i\chi}-1)\right]\,.
\end{equation}
This is one of the main results of our paper. 
It is valid in 2 and 3 dimensions; the dimension will only enter 
into the density of states, when
transforming the sum over $k$ into an energy integration via the
standard expression $N_D=m^{D/2} (2
\epsilon)^{D/2-1}/2\pi^{D-1}\hbar^{D}$ for $D=2,3$.  It should be noted,
that in a strictly two-dimensional system the low-energy scattering
amplitude vanishes $\sim -1/\log{\epsilon}$ and consequently the gap
equation (\ref{eq:gap-xi}) shows a logarithmic divergence for
$\epsilon\rightarrow 0$. However, for the more realistic situation of
a quasi-two-dimensional cloud (i.e., a three-dimensional trapped
atomic cloud strongly confined in one dimension), this singularity is
eliminated.  Results derived for the strictly two-dimensional
situation are still valid for the quasi-two-dimensional case, however
with the chemical potential $\mu$ shifted by the ground-state energy.

We now discuss some limiting cases in which analytical expressions can
be obtained. On the BEC side, $\mu<0$ and $\Delta\ll|\mu|$ leads to
$v_k^2\ll 1$ for all energies and allows to
expand the logarithm in Eq.~(\ref{eq:cgf-bcs}). The result is 
\begin{equation}
  \label{eq:cgf-beclimit}
  S(\chi)=-\frac{\bar N}{2}\left(e^{2i\chi}-1\right)\,,
\end{equation}
which corresponds to a Poissonian number statistics of \textit{pairs
  of atoms}. This supports the picture of strongly bound pairs in a
coherent state.  Note, that the factor of $2$ in the exponent leads to
exponentially growing cumulants in the fermion number, viz., $C_n=2^n
\bar N/2$. We therefore expect strong fluctuations. Remarkably, the
result (\ref{eq:cgf-beclimit}) is in accordance with the number statistics
of condensed bosons in a bin with volume much less that the total
volume, see the expression given after Eq.~(\ref{eq:cgf-binbosons})
that was derived using the canonical formalism. 
Since we are counting single fermions instead of bosons in the present
case, the counting field $\chi$ is replaced by $2\chi$, and there is a
prefactor $1/2$. This agreement is quite remarkable, 
since the starting point of our approach is the grand-canonical formalism.
It is an indication that using the BCS
wave function to describe the number fluctuations of a small subsystem
works surprisingly well.

On the BCS side the situation is quite different. Here
$\mu=\epsilon_F\gg\Delta$ and we obtain
\begin{equation}
  \label{eq:cgf-bcslimit}
  S(\chi)=-i\chi\bar N -  \pi\bar N  D 
\frac{\Delta}{4\epsilon_F}(|\cos(\chi)|-1)\,.
\end{equation}
We observe that the first term is dominant but contributes only to the
first cumulant. The fluctuations come from the second term in
Eq.~(\ref{eq:cgf-bcslimit}) which is smaller by a factor
$\Delta/\epsilon_F$. Furthermore, similar to the degenerate Fermi gas,
the odd cumulants $C_n$ for $n \ge 3$ vanish, which is again a
consequence of particle-hole symmetry.

Due to the constant density of states in 2D we can obtain an analytical
expression for the CGF for arbitrary $\mu$ and $\Delta$, which reads
\begin{widetext}
\begin{equation}
  S(\chi)=-\bar N\frac{\Delta}{\epsilon_F} 
  \left[\cos(\chi)\mathrm{atan}\left(\frac{2\epsilon_F}
   {\Delta}e^{i\chi}\right)-
    \mathrm{atan}\left(\frac{2\epsilon_F}{\Delta}\right)\right]
  -\frac{\bar N}{2} \frac{\mu}{\epsilon_F}
\ln\left[1+v_0^2(e^{2i\chi}-1)\right]\,.
\end{equation}
\end{widetext}
Here $v_0^2=(1+\mu/\sqrt{\mu^2+\Delta^2})/2$ is the BCS
coherence function for $k=0$.  While we do not have an 
analytical expression in 3D, we expect a similar behavior as in 2D. This
will be corroborated later by comparing the numerically obtained
cumulants in 3D to the 2D case.
\begin{figure}[htbp]
  \centering
  \includegraphics[width=8cm]{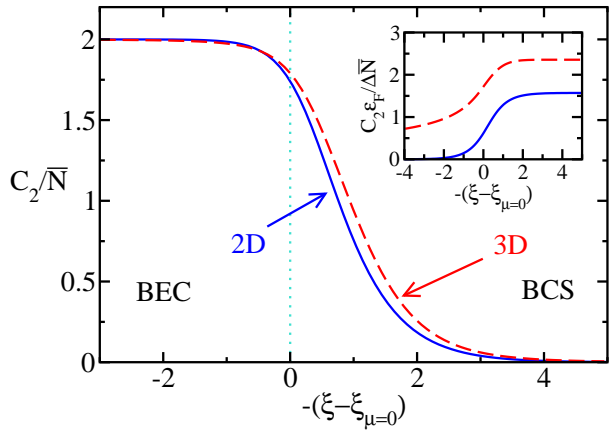}
  \caption{(Color online) Second cumulant $C_2$ as a function 
    of $\xi=1/k_F a$.  For the
    strongly confined, quasi-two-dimensional system (2D), the Fermi
    vector can be approximated by the inverse of the ground
    state size: $k_F = \pi/\ell_0$.  The dotted line corresponds to
    $\xi_{\mu=0}$. The inset shows $C_2$ normalized to
    $\Delta/\epsilon_F$.}
  \label{fig:c2}
\end{figure}

\begin{figure}[htbp]
  \centering
  \includegraphics[width=8cm]{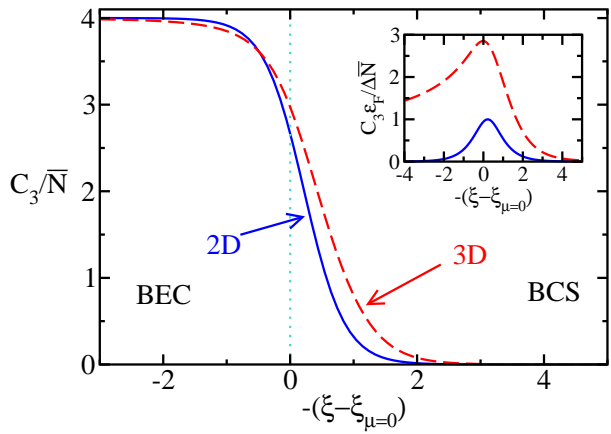}
  \caption{(Color online) Third cumulant $C_3$ as a function 
    of $\xi$ as defined in 
    Fig.~\ref{fig:c2}.  The dotted line corresponds to $\xi_{\mu=0}$.
    The inset shows $C_3$ normalized to $\Delta/\epsilon_F$.  }
  \label{fig:c3}
\end{figure}

In Fig.~\ref{fig:c2} we show how the density noise changes through the
BEC-BCS crossover. Going from the BEC to the BCS regime strongly
suppresses the fluctuations in agreement with our previous
discussion. The inset shows the fluctuations normalized to
$\Delta/\epsilon_F$; they approach a constant value in the BCS limit.
Another important information gained from a noise measurement is the
order parameter, which can be extracted from a measurement of $C_2$ in
the BCS limit, Eq.~(\ref{eq:cgf-bcslimit}), as $\Delta/\epsilon_F =
4C_2/\pi\bar N D$.  Figure~\ref{fig:c3} shows the third cumulant. The
global behavior is rather similar to the second cumulant, i.e. $C_3$
is strongly reduced going from the BEC limit to the BCS
limit. However, the behavior of $C_3$ normalized to
$\Delta/\epsilon_F$ shown in the inset is qualitatively different of
that shown in the inset of Fig.~\ref{fig:c2} since $C_3$ vanishes also
in the BCS limit. Note also that $C_3$ vanishes faster in the 2D case
than in the 3D case. To understand the behavior of the third cumulant
in more detail we recall that $C_3$ is related to the skewness of the
distribution, i.e. $C_3$ is a measure of the difference between
positive and negative fluctuations. To see this, we note that an
elementary binomial event has the property
$\ln(1+v_k^2(\exp(i\chi)-1))=i\chi+\ln(1+u_k^2(\exp(-i\chi)-1))$.
Using this property and the particle-hole symmetry of $v_k^2$ in the
BCS limit, we can rewrite the CGF in the BCS regime as $i\chi\bar N +
N_0 \int d\epsilon \ln[1+v_k^2u_k^2(\cos(2\chi)-1)]$. Here we have
used that the density of states close to $\epsilon_F$ can be
approximated by a constant $N_0$. The second term is even in $\chi$
and therefore only contributes to even cumulants, whereas all odd
cumulants for $n\ge 3$ vanish. The difference between 2D and 3D seen
in Fig.~\ref{fig:c3} is caused by the (small) energy dependence of the
3D density of states, which is absent in 2D.

\section{Finite temperatures}
\label{sec:finiteT}

We would now like to discuss the effect of finite temperatures in a
qualitative way. The excitations in the BCS theory are fermionic
quasiparticles.  This is a good approximation on the BCS side of the
transition. The fluctuations (measured by the second cumulant $C_2$)
will be reduced with increasing temperature, since they approach the
linearly increasing $C_2$ of the free Fermi gas that is lower at the
critical temperature than $C_2$ at $T=0$, 
$C_2^{\text{free}}(T_C^{\mathrm{BCS}}) < C_2(0)$. At even higher temperatures
$T\geq T_F$, $C_2$ reaches the classical Poisson value $\bar N$. The
situation on the BEC side of the crossover will be very different.
According to our result the statistics is a Poissonian distribution of
molecules and, hence, $C_2$ is doubled in comparison to the classical
value for the atomic gas \cite{Note}.  Upon increasing the
temperature, the main effect on the fluctuations will be a breaking up
of molecules, which will reduce the second cumulant $C_2$ above a
dissociation temperature to finally reach the value for the classical
gas. Thus, we expect quite a different temperature dependence of the
number fluctuations on the BEC or the BCS side of the transition.

\section{Conclusion}

In conclusion, we have calculated the full counting statistics of
number densities in an ultracold fermionic atomic cloud with
attractive interactions. The number statistics in the vicinity of the
BEC-BCS crossover displays interesting features which reveal the
nature of the many-body ground state. Poissonian fluctuations of a
molecular condensate on the BEC side are strongly suppressed on the
BCS side. The size of the fluctuations in the BCS limit is a direct
measure of the pairing potential.  We have also discussed the BEC-BCS
crossover of the third cumulant and the temperature dependence of the
second cumulant. These quantities can be accessed experimentally and
provide additional information on the many-body ground state in the
crossover regime.  The concept of counting statistics in ultracold
gases opens interesting possibilities to study the interplay between
coherence and correlation in quantum many-body systems.

\begin{acknowledgments}
  We would like to thank A. Lamacraft for discussions. This work was
  financially supported by the Swiss NSF, the NCCR Nanoscience, and
  the European Science Foundation (QUDEDIS network). This research was
  supported in part by the National Science Foundation under Grant
  No. NSF PHY05-51164.
\end{acknowledgments}


\begin{thebibliography}{99}
\bibitem{T_BEC} The expression for $T^{\mathrm{BEC}}_C$ is calculated for a
  three-dimensional gas confined by rigid walls. $\zeta$  is the Riemann zeta
  function.
\bibitem{Note} Note that here the fluctuations of atoms are discussed. This  
 should not be confused with the fluctuations of the molecule number.
\end{thebibliography}
\end{document}